\documentclass[sn-mathphys-num]{sn-jnl}% Math and Physical Sciences Numbered Reference Style 

\usepackage{graphicx}%
\usepackage{multirow}%
\usepackage{amsmath,amssymb,amsfonts}%
\usepackage{amsthm}%
\usepackage{mathrsfs}%
\usepackage[title]{appendix}%
\usepackage{xcolor}%
\usepackage{textcomp}%
\usepackage{manyfoot}%
\usepackage{booktabs}%
\usepackage{algorithm}%
\usepackage{algorithmicx}%
\usepackage{algpseudocode}%
\usepackage{listings}%

\usepackage{csquotes}
\usepackage{lmodern}
\usepackage{anyfontsize}

\begin{document}

\title[Software Engineering Podcasts]{Software Engineering Podcasts: An Empirical Study of Their Potential as a Research Resource}

\author*[1]{\fnm{Marvin} \sur{Wyrich}}\email{wyrich@cs.uni-saarland.de}

\author[2]{\fnm{Marcos} \sur{Kalinowski}}\email{kalinowski@inf.puc-rio.br}

\author[3]{\fnm{Adolfo} \sur{Neto}}\email{adolfo@utfpr.edu.br}

\author[1]{\fnm{Sven} \sur{Apel}}\email{apel@cs.uni-saarland.de}

\affil[1]{\orgname{Saarland University}, \orgaddress{\country{Germany}}}

\affil[2]{\orgname{Pontifical Catholic University of Rio de Janeiro}, \orgaddress{\country{Brazil}}}

\affil[3]{\orgname{Universidade Tecnológica Federal do Paraná}, \orgaddress{\country{Brazil}}}

\abstract{Podcasts have become an increasingly popular medium for knowledge sharing within the software engineering (SE) community, offering insights into industry developments and the perspectives of professionals with different backgrounds. As this medium grows, it presents a potentially valuable resource not only for practitioners but also for researchers seeking to understand the evolving field. However, little is known about the actual content of SE podcasts or how they are perceived and used by researchers. This study systematically explores the SE podcast landscape, analyzing its content and surveying researchers to assess how podcasts can serve as a meaningful resource for advancing empirical software engineering research.}


\maketitle

\section{Introduction}\label{sec:intro}

On the way to work, while cooking, or to help them fall asleep, millions of people listen to podcasts.
Among them is Alex, a software engineering (SE) researcher who starts the day with a news recap, unwinds in the evening with interviews on AI trends, and occasionally listens to academic peers discussing their work.
Podcasts have quietly become part of the academic routine, offering a low-effort, high-reward way to stay informed, curious, and connected.
For example, the \emph{Software Engineering Radio} podcast, published by IEEE Software, has become one of the most popular in the community, with now more than 700 published episodes~\cite{Blumen:2022:SERadio}.
And it is not just about listening. 
More and more researchers are being invited to speak on podcasts, share their work, and reflect on their experiences~\cite{Persohn:2025:Dissemination}.
It is therefore reasonable to assume that SE researchers are relatively familiar with podcasts as a medium.

Although many SE researchers like Alex are familiar with podcasts as part of their daily routine, this familiarity has not yet translated into a clear way to leverage podcasts for addressing a major challenge: connecting research with real-world practice.
The SE research community still struggles to understand practitioner needs~\cite{Wyrich:2025:Silent}.
Direct access to practitioners is often limited---especially for early-career researchers without established industry contacts---and traditional approaches to bridging this gap can be slow or burdensome.
Recognizing this, the community has started inviting practitioners to contribute directly to academic venues, hoping to better align research with industry challenges.
Prominent examples include the JSS column \enquote{Dear Researchers}~\cite{Avgeriou:2024:DearResearchers}, where practitioners are invited to write a short article addressing SE researchers, as well as the ICSE Industry Challenge Track~\cite{icse2026_industry_challenge}, where practitioners are invited to present industry challenges at an academic conference.

While these initiatives reflect a genuine effort to foster dialogue, they still rely on practitioners taking time to enter academic spaces, adopt academic formats, and speak the language of research. But practitioners are already communicating---openly, informally, and at scale---on platforms and in formats they choose themselves.
It may not always be necessary to ask them to write or come to academic venues.
It may simply require \emph{listening}.

In fact, practitioners increasingly share their experiences and perspectives through informal channels such as blog posts and social media~\cite{Garousi:2020:BenefittingGrey}.
These self-initiated formats allow for more candid, timely, and context-rich communication than traditional academic venues.
In empirical software engineering, such grey literature has gained recognition as a valuable data source, particularly for exploring practitioner needs and practices~\cite{Kamei:2021:GreyCriticalReview}.
Podcasts, with their often long-form, conversational nature, offer similar potential---but unlike blogs or social media platforms, they remain largely unexplored in SE research.
This raises the question of whether we are overlooking a rich and accessible resource already used and produced by the very communities we aim to study.

Early mentions of SE podcasts in the scientific literature date back to 2007, when a paper in the ACM SIGSOFT Software Engineering Notes explained what a podcast is and pointed to existing SE podcasts that would offer \enquote{information on books on new technologies such as test-driven development, features about software inspections and reviews, or interviews with celebrities such as Kent Beck}~\cite{Rech:2007:SEPodcasts}. Since then, however, we found no more than a handful of studies that have used podcasts as a data source in SE research; either to motivate research questions or to answer them directly. 
Two recent examples from 2025 include a study of podcast interviews to identify challenges and motivating factors for girls in STEM~\cite{Novaes:2025:WomenPodcastStudy}, and another exploring the strategies and experiences of individuals returning to the workforce after a career break~\cite{Ryan:2025:PodcastBreak}.
As \citet{Ryan:2025:PodcastBreak} concluded, podcasts offer context-rich, publicly available data without the need for recruitment or data collection. At the same time, they come with challenges such as limited control over the conversation, ethical ambiguity, and the lack of follow-up questions.
Balancing these trade-offs, the authors argue that podcasts represent a valuable yet underused resource---especially for qualitative research---while also highlighting the need for clearer guidance on how to work with such data~\cite{Ryan:2025:PodcastBreak}.

Despite their potential, podcasts remain a rarely used resource in empirical software engineering research. Uncertainty about how to find relevant podcasts, assess their quality, or integrate them into established research workflows may explain their limited adoption.
Additional challenges known from other grey literature may also play a role here (e.g., lack of structured information and reliability concerns)~\cite{Kamei:2021:GreyCriticalReview}, while at the same time it is difficult to assess how well existing guidelines for working with grey literature apply to podcasts~\cite{Garousi:2019:GuidelinesGrey}.
These issues point to a broader gap: Although podcasts may offer accessible, content-rich insights into practitioner perspectives, we know little about the actual landscape of SE podcasts, how researchers perceive their value, and what barriers hold them back from making use of this resource.
To address this, we conducted a study consisting of two complementary components. 
First, we analyzed the current landscape of SE podcasts to explore their episode formats and thematic focus.
Second, we surveyed SE researchers to understand how they perceive podcasts as a resource for empirical research. 
Correspondingly, this work is guided by the following two research questions:

\begin{description}
    \item \textbf{RQ1:} What topics and formats characterize the current landscape of SE podcasts?
    \item \textbf{RQ2:} How do SE researchers perceive podcasts as a resource for empirical research? % delete empirical?
\end{description}

\section{Methodology}\label{sec:methods}

To answer the previously motivated research questions, our approach is twofold, consisting of a systematic podcast landscape analysis~(Section~\ref{methods-podcasts}) and a survey among researchers~(Section~\ref{methods-survey}). For transparency and reproducibility, we share our study artifacts online\footnote{Replication package: https://doi.org/10.5281/zenodo.18962642}. This includes a tabular overview of the sampled podcasts and their extracted attributes. It also includes the questionnaire and anonymized responses.

\subsection{Research Context}

A podcast is \enquote{a program (as of music or talk) made available in digital format for automatic download over the Internet}~\cite{mw:podcast}. In this study, we focus specifically on podcasts that feature spoken content related to software engineering, excluding music or purely entertainment-focused shows. Each podcast consists of episodes, and both podcasts and episodes include metadata such as titles, descriptions, and lengths, which are integral to our analysis.

Several platforms exist for hosting and/or distributing podcasts, including popular options such as Apple Podcasts, Google Podcasts, Amazon Music, Stitcher, and the option for self-hosting on personal websites. For the selection of podcasts, we chose \emph{Spotify}\footnote{https://www.spotify.com} as the primary platform, given its wide availability and comprehensive podcast catalog, which makes it a reasonable proxy for the broader podcast ecosystem.

It is important to note that the third author of this paper is the host of a software engineering podcast, bringing firsthand experience to the study.
His podcast is distributed on platforms such as Spotify.
However, none of the authors, including the third author, stand to gain personally or professionally from research on this specific topic, and we have no affiliation with Spotify.
Our design choices have been made independently and based solely on what we consider to be the most effective approach to answering our research questions.

\subsection{Podcast Landscape Analysis}\label{methods-podcasts}

The objective of this first part of our study is to find and analyze a representative sample of software engineering podcasts to better understand the diversity in content and form of such podcasts. 

\subsubsection{Podcast Selection}

We define software engineering podcasts as podcasts that deal thematically with the development of software systems. These may include discussions on topics related to the act of programming, design, testing, management, and maintenance of software systems, as well as discussions about the people who participate in the software process, their professional practices, aspirations, challenges, and skills.

It is opportune to define the term rather broadly to be able to identify a certain variety of podcasts and potentially also to find podcasts that are at the intersection to other disciplines. We consider fully automatic filtering, for example, based on terms that must appear in the title of the podcasts, to be too restrictive. One reason for this is that podcasts tend to use creative wordplay in their names, which are recognizable to a human, but not a keyword search.

We therefore define a set of inclusion criteria, the fulfillment of which is assessed by three authors of this study, and which must be considered fulfilled by all three for a podcast to be included for further analysis:

\begin{itemize}[\indent {}]
    \item[\textbf{I1~}] At least three out of the podcast's six most recent episodes explicitly refer to software engineering in their title or description, according to our definition of SE podcasts provided above.
    \item[\textbf{I2~}] The language of the podcast is English.
    \item[\textbf{I3~}] At least three episodes have been published.
    \item[\textbf{I4~}] At least one episode has been published since 2021.
\end{itemize}

We chose inclusion criterion I2 to ensure that our study results can be transparently understood by the international research community. Inclusion criterion I3 is intended to ensure that our results are not skewed, for example, by podcasts that have discontinued after only one or two episodes or merely serve as a storage location for a single audio file.
We further aim at drawing an up-to-date picture of the current podcast landscape, which is why we focus on podcasts that have been active within, at least, the past five years (I4).

We chose Spotify as a representative of a podcast provider and used the Spotify API to search for podcasts. We used the search term \enquote{software engineering} across the markets of the United States, Great Britain, Australia, and India---markets where the language of the target audience is likely to be English and which also have a good geographical distribution.\footnote{Note that this does not explicitly restrict the markets to these four, but merely serves as a parameter required for the search. As can be seen from our replication package, all the included podcasts are available in several dozen countries worldwide, including one of the four listed ones.} The search was conducted on September 19, 2024. After the removal of duplicates and automatic filtering according to inclusion criteria I3 and I4, 828 results remained. For each of these 828 podcasts, a manual decision was then made based on the inclusion criteria as to which ones would be considered for further analysis. This initially led to the inclusion of 224 podcasts. After a sanity check by the fourth author and an attempt to listen to each podcast at least once, eight more were excluded due to lack of access or contents that did not correspond to the podcast description. \textbf{We ended up with 216 English-language software engineering podcasts} for further analysis. Metadata and analyses of these podcasts were retrieved and conducted in June 2025.

\subsubsection{Data Analysis}

We manually categorized each included podcast, both in terms of its content and its format.
We then supplemented these data with metadata for each podcast, retrieved via the Spotify API.
The categorizations, together with the metadata, were used to characterize the current landscape of SE podcasts and thus answer our first research question.

\paragraph{Content Categorization} 
Our initial attempt at categorizing content was based on the research areas listed in the ICSE 2025 call for papers. This call defines nine research areas, each with bullet-pointed subtopics, which we assumed would broadly reflect the thematic landscape of software engineering.
In a first trial, we were able to assign a subset of podcasts to, at least, one of these areas. However, we soon observed that the more episodes we examined from a given podcast, the more likely it was that nearly all areas would be covered, especially in podcasts with a wide-ranging focus. The ICSE categories offer a fairly academic partitioning of research topics, which may work well for structuring scholarly work, but are less suitable for classifying the more fluid, overlapping discussions typically found in podcasts. This led to inconsistent assignments between authors and an overall low explanatory value of the classification.

As a result, we shifted to a higher-level, inductively derived classification that better aligned with our observations of the podcast landscape. Each podcast was assigned to one or more of the following three topical categories:
\begin{itemize}
    \item \emph{Technical \& Practical Knowledge} -- On how software engineering works,
    \item \emph{Industry \& Trends} -- On what is happening in the software world,
    \item \emph{Career \& Social Aspects} -- On the people behind software engineering.
\end{itemize}

This revised categorization yielded a more consistent and meaningful classification of podcast content.

\paragraph{Format Categorization}
In addition to content, we also examined the format of SE podcasts. Drawing on format classifications established in prior podcast research from other domains~\cite{Rogers:2019:Podcasting,Moore:2024:Podcasts,Rime:2022:Podcast}, we adapted and refined these to suit the SE context---for example, some categories, such as \emph{fictional storytelling}~\cite{Rime:2022:Podcast}, did not play a role in our context. Overall, the literature did not offer a consistent solution either, likely because an appropriate categorization depends on the specific purpose. In our case, we aim to give SE researchers an intuitively understandable idea of what formats to expect when using SE podcasts as a data source, without being overly specific. Therefore, after two rounds of refinement, we arrived at the following two categories:

\begin{itemize}
    \item \emph{Interview and Narrative-Driven Podcasts} -- These focus on conversations between hosts and guests or feature structured storytelling. Formats include expert interviews, co-host discussions, or panel episodes that aim to provide engaging insights, experiences, or opinions.
    \item \emph{Monologues and Personal Journaling} -- These involve a single host sharing thoughts, experiences, or expertise in a reflective or informative manner, without external guests. Formats range from structured topic exploration to spontaneous, journal-like recordings.
\end{itemize}

A podcast was assigned to both categories if, at least, one episode matched each respective format. Additionally, we took notes on particularly distinctive formats or content, allowing us to revisit notable cases during analysis.

\paragraph{Metadata Analysis}

The content and format categorizations were supplemented by metadata retrieved via the Spotify API in June 2025. For each podcast, we collected: podcast ID, publisher, description, total number of episodes,  date and duration of the latest episode, podcast image URL, and a list of countries in which the podcast is available.

\subsection{Researcher Survey on SE Podcasts}\label{methods-survey}

The objective of this survey is to analyze SE podcasts with the purpose of characterizing their current usage and perceived value, as well as identifying barriers to their use and gathering recommendations, from the perspective of software engineering researchers in the context of conducting software engineering research.

To ensure scientific rigor, we follow \citeauthor{Kasunic:2005:Survey}’s seven-step guideline for survey design~\cite{Kasunic:2005:Survey}, complemented by \citeauthor{Linaker:2015:Survey}'s more recent annotations~\cite{Linaker:2015:Survey}. While a distinction is made between qualitative and quantitative surveys, we place greater emphasis on the latter, as our study primarily collects quantitative data through a questionnaire, which is subsequently analyzed statistically. However, a few qualitatively oriented questions aim to \enquote{find most, if not all, the possible values of [a] characteristic in the population}~\cite{Melegati:2024:QualitativeSurveys}.

\subsubsection{Sampling Strategy}
\label{sec:survey_sampling}

Our target population consists of adults who are currently or have previously conducted research in the field of software engineering and therefore have a connection to this research community. We use convenience sampling to distribute our survey. To reach as many participants as possible, we employ multiple distribution channels in parallel, including social networks and email lists. Additionally, we ask participants to forward the questionnaire to qualified colleagues.

\subsubsection{Questionnaire Design}

Two key considerations guided the development of the questionnaire. First, we aimed to ensure that participants would not need more than 15 minutes to complete it. Second, we designed the questions to be accessible and relevant even to members of the target group who have not had any direct experience with SE podcasts.

We divided the overarching objective into three subgoals, each associated with specific questions. These subgoals also formed the structure of the first three sections of the questionnaire. The \textbf{first section} includes four questions aimed at learning about the participant’s overall \emph{points of contact} with podcasts. The \textbf{second section} contains five questions aimed at understanding the \emph{value} that the participant sees in using SE podcasts for their work. The \textbf{third section} consists of three questions aimed at identifying \emph{barriers} that the participant sees in using SE podcasts for their work. A \textbf{fourth and final section} collects \emph{demographic characteristics} of the sample: primary role, number of years involved in SE research, age, and the region in which the respondent is primarily based. The questionnaire ends with an open-ended question inviting participants to share any additional thoughts they may have.

Of the 17 total questions, four are open-ended, meaning that \enquote{respondents create their own answers to the question in their own words}~\cite{Kasunic:2005:Survey} without being presented with predefined options. The remaining questions offer either multiple-choice responses or ask participants to position themselves on a scale. One question requires participants to rank the answer options from most relevant to least relevant (\enquote{How do SE podcasts compare to other sources of information you use for research?}).

The complete questionnaire, including all questions and response options, is available in the replication package.

\subsubsection{Pilot Test and Data Collection}

Three individuals from the target population were asked to complete the questionnaire (1 professor, 1 PhD student, 1 industry researcher). They received the same information as the later participants of the official survey, including the informed consent form. However, the pilot participants were additionally asked to tell us about any difficulties in understanding specific questions or other irregularities. Their feedback led to several minor revisions.
We updated the informed consent form and refined the definition of SE podcasts at the beginning of the study.
We also added a note encouraging participants to share podcast examples even if they were not entirely certain they qualified as SE podcasts.
In addition, we split one question into two after a pilot participant indicated they would respond differently to the individual parts, and we reworded one response option to improve clarity. The estimated survey duration of 10 to 15 minutes was met by all test participants.

The survey was online for 21 days. At the beginning, halfway through, and at the end, we reminded potential participants about it. In total, 83 participants agreed to the informed consent form. Of those, 53 went through all pages of the survey and completed the final section on demographic characteristics. Note that submission was not required, and we also included partial responses from participants who did not complete the entire survey.

\subsubsection{Data Analysis}

Our overall data filtering strategy was twofold: First, we removed all cases in which participants had clicked through the survey without providing any responses. Then, for each individual question, we performed sanity checks where appropriate and analyzed the responses we had available. As a result, the number of valid answers may vary across questions.

Since this is an initial exploratory investigation, we did not formulate hypotheses in advance.
Accordingly, our analysis is largely descriptive and exploratory in nature.
In Section~\ref{sec:results}, we present what we observe in the data. We deliberately separate this from the discussion in Section~\ref{sec:discussion}, where we explicitly answer our second research question on how SE researchers perceive podcasts as a resource for empirical research and reflect on the implications of our findings for the research community.
Of course, every decision about how to present results involves a degree of filtering. To support transparency and enable further exploration, we therefore share the raw survey data in our replication package.

\section{Results}\label{sec:results}

In this section, we present the results of our study in two parts: the podcast landscape analysis (Section~\ref{sec:pla}) and the researcher survey on SE podcasts (Section~\ref{sec:surveyresults}).

\subsection{Podcast Landscape Analysis}
\label{sec:pla}

To explore the landscape of software engineering podcasts and the variety of topics and formats they offer, we analyzed a dataset of 216 podcasts. All results presented here are based on data collected up to June 2025.

\subsubsection{Podcast Metadata}

\begin{figure}[h]
    \centering
    \includegraphics[width=1\linewidth]{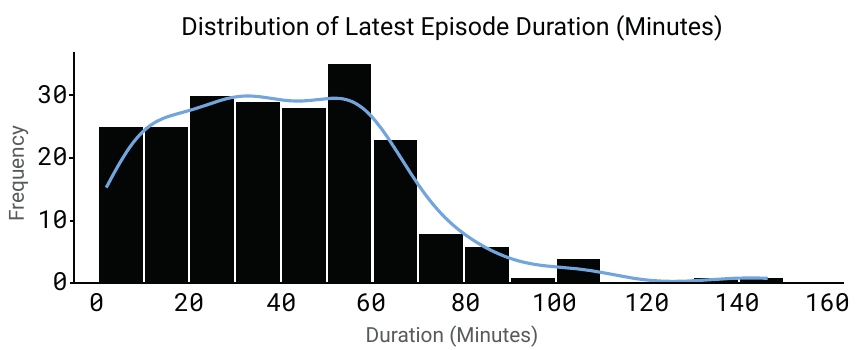}
    \caption{Histogram of the durations of the most recently published episode for each podcast in our dataset, aggregated in 10-minute bins.}
    \label{fig:episode_duration}
\end{figure}

The 216 podcasts in our dataset show substantial variation in both the number and duration of episodes. 
The median number of episodes per podcast is 34 (M = 126, SD = 270, range 3--2,093).
Considering the most recent episode of each podcast, the median duration is 39 minutes (M = 40, SD = 25, range 2.2--146 minutes).
Podcasts with particularly short episodes sometimes advertise this explicitly. 
For example, \emph{snackableCast -- Software Development \& Engineering Culture} promotes itself with \enquote{Short episodes. Real stories. Practical wisdom.}
Similarly, the podcast \emph{Around IT in 256 seconds} makes its format clear already in the title, noting in its description that episodes are \enquote{never longer than 4 minutes and 16 seconds} and targets listeners \enquote{who want to combat FOMO, while brushing [their] teeth}.
A plot of the latest episode duration is provided in Figure~\ref{fig:episode_duration}, showing that the majority of podcasts have episodes between 20 and 60 minutes in length.
Within this range falls, for example, the podcast \emph{Software Engineering Radio}---mentioned in the introduction and perhaps familiar to some readers---which, with its 680 episodes, typically runs for about 50 minutes each.

Nearly half of the podcasts ($\sim$47\%) can be considered still active, having released, at least, one episode in the first half of 2025.
The podcasts are almost entirely published by distinct publishers, with only three exact name matches where a publisher is associated with two podcasts.
Since podcast providers can generally choose their names freely---and individuals may also co-host across different shows---this is not a perfect indicator of unique entities. Still, the overall picture suggests that the podcast landscape is characterized by a number of providers almost as large as the number of podcasts themselves.

\subsubsection{Podcast Topics}

\begin{figure}
    \centering
    \includegraphics[width=1\linewidth]{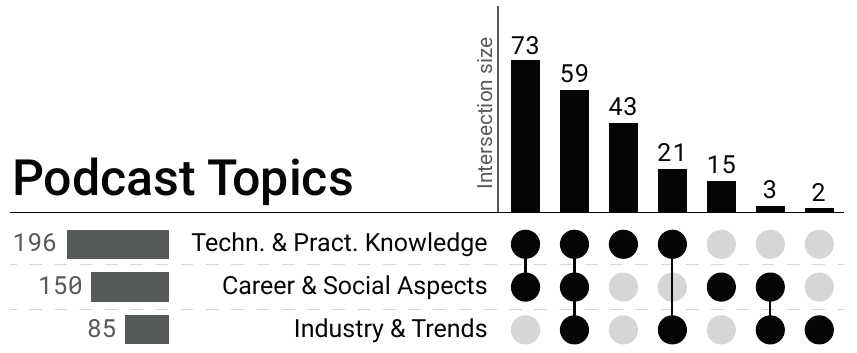}
    \caption{Distribution of SE podcast topics, showing how the 216 podcasts are assigned to one or multiple of three topical categories; intersection size denotes an absolute number of podcasts.}
    \label{fig:topics}
\end{figure}

From the UpSet plot in Figure~\ref{fig:topics}, we see how many of the 216 podcasts were labeled as \emph{technical \& practical knowledge}, \emph{career \& social aspects}, \emph{industry \& trends}, or combinations thereof.
Unsurprisingly, technical topics appear most frequently, including in the largest set---podcasts that combine technical themes with career- and social-aspect content (73 podcasts fall into exactly these two categories).
The technical spectrum ranges from highly specialized shows focusing on particular languages and platforms (e.g., \emph{Talk Python To Me}, \emph{.NET Rocks!}, \emph{AWS Developers Podcast}) to broad, general-purpose SE podcasts that let current events shape their coverage (e.g., \emph{Software Engineering Daily}, or \emph{The Changelog: Software Development, Open Source}).
However, pure trend podcasts that provide regular insights into what is happening in the software industry but do not delve into technical details or link it to career aspects are less common.

Social aspects are also featured prominently, appearing in 150 of the 216 podcasts.
In 15 cases, they were the only topic category, centering exclusively on the people behind software development. A well-known example is \emph{Soft Skills Engineering}, described as \enquote{a weekly advice podcast for software developers about the non-technical stuff that goes into being a great software developer,} now with more than 450 episodes.
Many podcasts in this category address interpersonal communication and collaboration with other IT professionals, often shaped by the hosts' distinct perspectives.
In \emph{Spilling the ITea}, for instance, the three hosts describe it as \enquote{a podcast where we, Agnés, Maria and Emma, debug our experiences as women in tech and IT, specifically in Computer Science and Software Engineering,} deliberately foregrounding women's perspectives.
Other podcasts focus on personal and professional growth, such as the interview-based \emph{Dev Leader Podcast}, which aims to highlight \enquote{how diverse the field of software engineering is and how everyone has their own powerful story to share about their own career development.}

\subsubsection{Podcast Formats}

\begin{figure}
    \centering
    \includegraphics[width=1\linewidth]{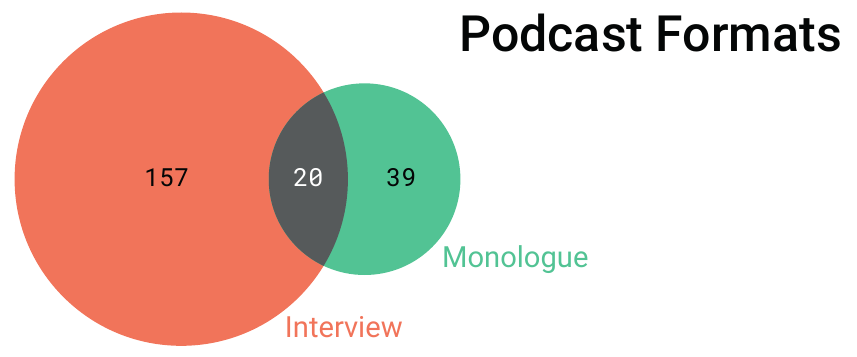}
    \caption{Distribution of SE podcast formats, illustrating how the 216 podcasts fall into the two format categories and their overlap.}
    \label{fig:formats}
\end{figure}

Regarding podcast format, there is a clear majority of shows in which at least two people speak: 157 out of 216 podcasts exclusively use this format (see Figure~\ref{fig:formats}). 
This includes, firstly, interviews with rotating guests, such as in the podcasts \emph{CodeNewbie} (\enquote{Stories and interviews from people on their coding journey}) or \emph{How AI is Built} (\enquote{Real engineers. Real deployments. Zero hype. We interview the top engineers who actually put AI in production}).
Secondly, it also includes discussions between two or more recurring hosts without necessarily bringing in external guests; the typical format here features two regular hosts.
One of the few exceptions is \emph{Testing Peers}, \enquote{a community-driven initiative built by testers, for testers,} in which three or more people often discuss topics together.

In contrast, 39 podcasts were categorized purely as monologues, in which a single host shares their thoughts, experiences, or expertise.
These are often technical or explanatory in nature, such as \emph{Learn System Design}, \enquote{a bi-weekly podcast hosted by a senior engineer named Ben Kitchell that takes a deep dive into learning about technical system design,} or \emph{The Backend Engineering Show with Hussein Nasser}, where the host \enquote{discuss[es] all sorts of software engineering technologies and news with specific focus on the backend.}

Only 20 podcasts exhibited a mix of both formats, with the number of speakers depending on the episode.
Overall, all SE podcasts in our sample can be classified using this simple format categorization and fit in very well.
Naturally, some podcasts combine content and format in ways that make them stand out.
Examples include \emph{Software Architecture Book Club}, which goes \enquote{chapter-by-chapter through books about software architecture,} and \emph{System Design}, where the hosts simulate technical job interviews.

\subsection{Researcher Survey on SE Podcasts}
\label{sec:surveyresults}

A total of 83 participants agreed to the consent form, of whom 67 provided at least one survey response.
Among these, 53 participants completed the entire survey, with a median time of around 7 minutes.

\subsubsection{Demographics}

\begin{figure}
    \centering
    \includegraphics[width=1\linewidth]{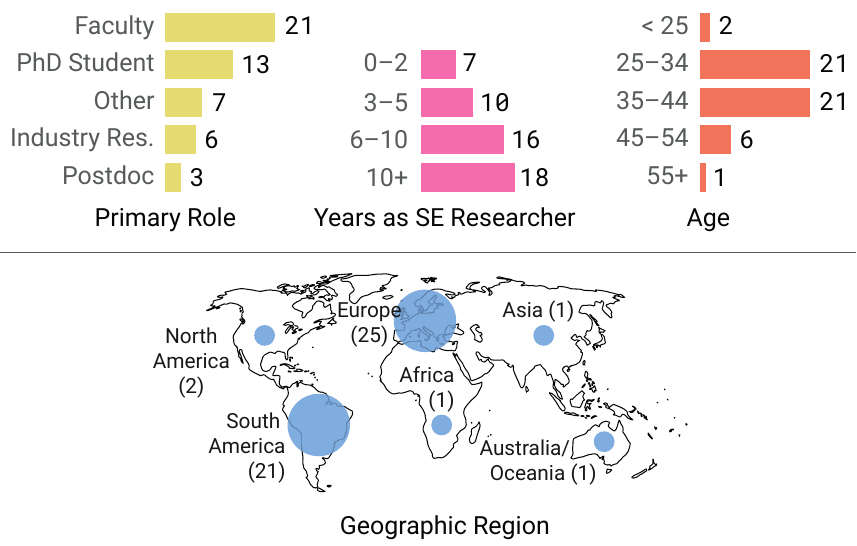}
    \caption{Demographic characteristics of survey participants. The figure summarizes respondents’ primary professional role, experience in software engineering research (in years), age distribution, and primary geographic region. All questions were optional; therefore, the number of responses varies across panels. One \enquote{prefer not to say} response for primary role was excluded from the visualization.}
    \label{fig:demographics}
\end{figure}

Figure~\ref{fig:demographics} summarizes the demographic characteristics of our survey participants.
The majority identified as faculty members or professors (21 responses) and PhD students (13). In addition, a smaller number of industry researchers (6) and postdoctoral researchers (3) participated, as well as seven respondents who selected the \enquote{other} option and identified themselves, for example, as \enquote{graduate students} or simply as \enquote{researchers.}

Regarding age, most participants reported being between 25--34 or 35--44 years old.
In terms of experience in software engineering research, the sample can be considered relatively experienced: 18 respondents reported more than ten years of research experience, whereas only seven indicated fewer than three years.
The descriptive data suggest a plausible association between the generally extensive research experience and the predominance of faculty members as the primary role.

While primary role, years of experience as a software engineering researcher, and age distribution reflect a diverse sample, the geographic distribution shows a strong concentration of respondents based in Europe and South America.
This is likely due to a network effect related to the authors of this study and is discussed further in the limitations (Section~\ref{sec:limitations}).

\subsubsection{Points of Contact}

\begin{figure}
    \centering
    \includegraphics[width=1\linewidth]{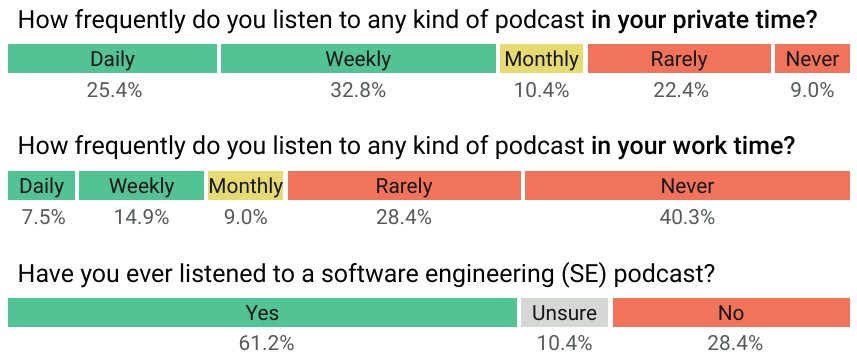}
    \caption{Respondents’ podcast listening frequency in private and work contexts, as well as prior exposure to software engineering podcasts ($n=67$ for all three measures).}
    \label{fig:points_of_contact}
\end{figure}

To gain a better understanding of our respondents' familiarity with podcasts in general and SE podcasts in particular, we first asked three questions about listening frequency (see Figure~\ref{fig:points_of_contact}).
It is noticeable that a large proportion of 68.7\% listen to podcasts, at least, monthly in their private time; about a quarter of our respondents even do so daily.
Regarding the question of how often they listen to podcasts during work time, the picture is reversed: about 68.7\% rarely or never listen to podcasts during work hours.
The six participants (9\%) who indicated that they never listen to podcasts in their private time also reported never listening during work time.
The five participants (7.5\%) who listen to podcasts daily during work hours also listen daily in their private time.
When asked whether they have ever listened to a software engineering podcast, 41 participants (61.2\%) answered yes, 19 (28.4\%) answered no, and 7 (10.4\%) were unsure.

A fourth question invited participants to list SE podcasts they had listened to in the past, with the option to include examples even if they were not completely certain these qualified as SE podcasts.
A total of 83 participants provided answers, mentioning 84 different podcasts.
With 18 mentions, the most frequently cited podcast was \emph{Fronteiras da Engenharia de Software}, an active podcast by Adolfo Neto (the third author), in which researchers from software engineering are interviewed about their work.
Other podcasts mentioned by at least four respondents include \emph{Software Engineering Radio}, \emph{Hipsters Ponto Tech}, \emph{Software Engineering Daily}, \emph{The Pragmatic Engineer}, and \emph{The Changelog}.
Among the English-language podcasts, only a handful were mentioned that were not already included in our sample in the podcast landscape analysis (Section~\ref{sec:pla}).
An ad hoc exploration of these podcasts shows that they were not manually excluded by us but were simply not returned by the Spotify API for our search term.
For example, \emph{The Haskell Interlude}, \enquote{where the five co-hosts [...] chat with Haskell guests!}, apparently does not mention \enquote{software engineering} in either the podcast description or episode descriptions and, we suspect, was therefore not found.
We return to this point in the discussion of limitations (Section~\ref{sec:limitations}).

\subsubsection{Perceived Value of SE Podcasts}

A key goal of this survey is to learn more about the perceived \emph{value} of podcasts as a research resource.
We asked four specific questions to this end.

% VA01: Software engineering (SE) podcasts are relevant, either to my current research or in future research.
% strongly disagree -- strongly agree

% VA02: What aspects of SE podcasts do you find valuable for your research, or think could be valuable in the future?
% checkbox choosing from options + "other" freetext

When asked whether SE podcasts are relevant to their current or future research, 7 respondents (11.5\%) strongly agreed, 22 (36.1\%) agreed, 23 (37.7\%) took a neutral position, 4 (6.6\%) disagreed, and 5 (11.5\%) strongly disagreed.
Accordingly, for roughly every second respondent, SE podcasts are either already relevant to their research or are seen as having potential relevance for future research.
For only about one in seven respondents, SE podcasts play no role in their own research.

To better interpret these numbers, we asked a follow-up question about aspects that make SE podcasts valuable for their research, either now or in the future.
We provided 5 response options with the possibility to select none or multiple options: learning about new research (43 responses), industry insights (38 responses), expert opinions (43 responses), and inspiration for new research questions (32 responses).
The fifth option, \enquote{Other (please specify),} was selected by 6 participants.
The median number of options selected by participants was 3.
The free-text responses highlight additional perceived value in SE podcasts, including the quality of discourse and interaction between hosts, insights into well-known products or offerings, and curiosity-driven engagement with specific topics or individuals.
One respondent also emphasized their usefulness as potential sources of empirical data.

% VA03: Can you recall a specific SE podcast episode that influenced your research or thinking? If so, which one and how?
% freetext

When asked whether participants remembered a specific SE podcast episode that influenced their research or thinking, and if so why, we received several insightful free-text responses.
One group of responses emphasized that certain podcasts lowered the barrier to entering the SE research field or to transitioning from SE research into a related area.
For example, respondents noted that \enquote{\emph{This IS Research} [\ldots] made me less anxious to enter a new field}, that an episode of \emph{Hipsters.tech} was their first time encounter with software architecture concepts during a transition into IT, and that episodes in the \emph{Future of Coding} (now: \emph{Feeling of Computing}) provided concrete impulses to a programming language researcher for thinking about certain SE concepts.
Another respondent explicitly attributed their decision to study SE to an episode on software architecture recovery from \emph{Fronteiras da Engenharia de Software}.\\
A second group of responses highlighted podcasts as sources of inspiration and concrete input for ongoing research.
One participant described episodes in which professors discussed current research topics as motivating for their own work, and that an episode of \emph{Elixir em Foco} helped them decide between Elixir and Clojure when selecting a functional programming language in the design phase of a study.\\
Finally, two responses praised the opportunity podcasts provide to engage deeply with complex topics without requiring extensive prior experience.
In particular, respondents emphasized the \enquote{ability [of certain hosts] to introduce complex concepts in an accessible way}, which made in-depth engagement possible even for newcomers.

% -------

% VA04: How do SE podcasts compare to other sources of information you use for research?
% ranking 1-6

\begin{figure}
    \centering
    \includegraphics[width=1\linewidth]{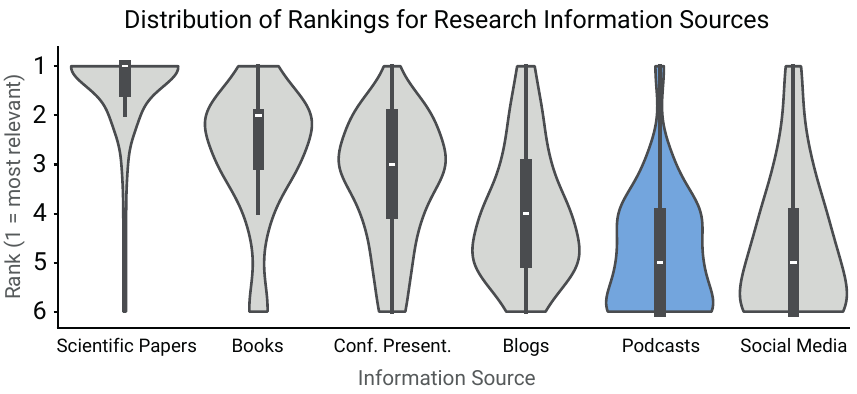}
    \caption{Violin plots show how participants ranked six sources of information for their research work, from most relevant on top to least relevant on the bottom ($n=59$).}
    \label{fig:sources_relevance}
\end{figure}

As part of our investigation into the potential value of podcasts, it was also important to understand how useful researchers perceive podcasts compared to other sources of knowledge for their work.
We presented six information sources for their research work and asked participants to rank them from most relevant to least relevant.
Figure~\ref{fig:sources_relevance} shows the distribution of rankings for each of the six information sources.
Unsurprisingly, scientific papers were ranked as the most useful source, followed by books and conference presentations.
By comparison, blog posts were perceived as less relevant, while podcasts and social media networks shared the lowest rank.

\subsubsection{Perceived Barriers to Using SE Podcasts}

The goal of this part of the survey is to identify the barriers that limit the use of podcasts as a research resource.
We first asked participants what challenges they see in using SE podcasts as a research resource.
We provided six answer options, of which none or several could be selected: lack of credibility (15 responses), hard to find and/or reference (26 responses), too time-consuming (28 responses), not research-focused (14 responses), not enough high-quality content (9 responses), and \emph{other} with the option to fill in a free text field (9 responses).
The median number of options selected by participants was 2.
Five participants provided additional information on the \emph{other} option.
They cited challenges such as \enquote{informality}, \enquote{subjectivity}, \enquote{the difficulty in explaining ideas (sometimes complex), without the support of visual or written content}, lack of clarity whether what is said \enquote{are findings grounded in research or [\ldots] just opinions and ideas}, \enquote{lot of noise (other data)}, and the unlikely prospect of \enquote{getting reviewers accept [podcasts] as reliable source of information}.

Then we turned the question around and asked our participants what would make SE podcasts more useful for their research or how they think they could be effectively used as a research resource---even if they do not currently use them themselves.
From the 25 responses received, we identified four overarching themes.

A major concern regarding the use of podcasts for scientific purposes relates to their credibility and doubts about whether reviewers would accept them as a legitimate resource.
Some respondents acknowledged their potential usefulness but emphasized that podcasts should be used alongside other sources, \enquote{but not the sole reference.}
Others primarily viewed podcasts as a channel for science communication rather than as a primary source of information for conducting research.
A frequently mentioned suggestion to enhance credibility was that podcast hosts should provide \enquote{sufficient references} and \enquote{sources for the topics they discuss} in episode descriptions.
Respondents also noted that, \enquote{as any other grey literature source, [podcasts] need an additional evaluation to consider what is said and by whom.}

A second commonly raised aspect concerned improved visibility and discoverability of podcasts.
Some respondents called for \enquote{a high-quality dataset of SE podcasts with reliable means of automatic analysis,} while others suggested \enquote{a single web page or something where all/most podcasts can be found,} ideally searchable by keyword, length, topic, or speaker background.
Another proposal envisioned \enquote{a globally established platform} for SE podcasts, comparable to Google Scholar, arXiv, or Zenodo, but tailored specifically to software engineering.
Overall, the responses indicate a clear desire to make podcasts easier to find and to introduce an additional layer that supports their use for research purposes.

A third theme was the wish for citable transcripts, ideally allowing specific statements to be linked to exact timestamps within an episode.
Transcripts would also make podcasts more accessible overall, as some individuals find it easier---or only possible---to process information in written form.

Finally, a fourth theme emphasized content-related expectations.
Respondents suggested that SE podcasts would be more useful if they focused on current industrial practices, tools, and programming languages, and if they provided expert insights, practical tips, and guidance for conducting research.
In particular, podcasts were seen as well suited for sharing informal, experience-based knowledge and real-world perspectives that often do not appear in formal publications.

\section{Related Work}\label{sec:relatedwork}

To contextualize our findings, we situate this study in prior research on grey literature in software engineering and on podcasts as a scholarly resource.

\subsection{Grey Literature in Software Engineering}
Grey literature (GL) has long been recognized as a potentially valuable complement to formal academic sources in software engineering.
Early multivocal literature reviews (MLRs), such as a technical debt study by \citet{Tom:2013:TechDebtMultiVocal}, already demonstrated the feasibility of systematically incorporating online, non-academic sources (e.g., blogs, wikis, websites).
Subsequent work further established methodological foundations and use cases for GL-inclusive research.
For example, \citet{Raulamo-Jurvanen:2017:ChoosingTestAuto} conducted a review on test automation tools based purely on grey sources, while \citet{Bogner:2021:Industry} combined interviews with the analysis of 295 practitioner resources, illustrating the growing methodological maturity of GL research in SE.

A key turning point was the call by \citet{Garousi:2016:NeedMultivocal} to increase awareness of multivocal literature reviews.
Their work argued that excluding grey literature can lead to substantial blind spots and may even steer research directions incorrectly.
At the same time, they noted that GL evidence is often opinion- or experience-based rather than grounded in systematic data collection, highlighting the need for methodological care.
Building on this, \citet{Garousi:2019:GuidelinesGrey} proposed comprehensive guidelines for conducting MLRs in SE.
Importantly, they criticized a double standard in the community: practitioner interviews reported in academic papers are widely accepted, whereas practitioner-authored grey sources (e.g., blogs) are often dismissed as unscientific.
Their guidelines aim to enable more rigorous and transparent use of such materials.

Empirical studies confirm both the prevalence and the challenges of using grey literature.
\citet{Yasin:2020:UsingGrey} found that 76\% of SE systematic literature reviews include, at least, one grey source, although GL accounts for only about 9\% of all cited primary studies. Similarly, \citet{Kamei:2021:EvidenceGreyLit} showed that GL often provides evidence not present in traditional literature and is frequently used for explanations and recommendations to solve problems.
Notably, the most common GL types reported are practitioner-produced artifacts such as blog posts, slide decks, and project descriptions---podcasts are not represented in these analyses.
Yet, in our sample, 68\% agreed or strongly agreed that SE podcasts can be considered a form of grey literature, suggesting that positioning podcasts within the GL discourse is appropriate.

Perception studies among SE researchers reveal persistent skepticism.
In a survey of 76 researchers, \citet{Kamei:2020:GreyLitSurveyBra} found that, while GL is valued for understanding practical problems and complementing research findings, over half of respondents avoid citing it due to concerns about scientific credibility.
The most frequently reported challenges were lack of reliability, lack of scientific value, and difficulties in searching and structuring GL sources.
These concerns closely mirror those raised by participants in our own survey.

Recent methodological discussions further refine when and how GL should be used.
\citet{Kitchenham:2022:ShouldSESecInclude} emphasize that the unit of analysis in systematic reviews is the primary study rather than the publication venue.
Consequently, non-traditional sources may be admissible if they report rigorous empirical work and remain publicly accessible long term.
Complementing this perspective, \citet{Williams:2017:BlogArticlesAsEvidence} specifically examine practitioner blog articles as an empirical evidence source.
Their findings suggest that blog content can provide valuable practitioner-generated evidence if it meets criteria such as rigor, relevance, clarity, and experience grounding.
This work aligns with our narrative by making a medium-specific argument---similar to our focus on podcasts---rather than treating grey literature only at an aggregate level.

Taken together, prior work establishes both the promise and the methodological challenges of incorporating grey literature into SE research.
However, existing studies focus predominantly on written practitioner artifacts, with little to no explicit attention to podcasts as a distinct and increasingly popular medium.

\subsection{Podcasts as Empirical Resource}

The idea of podcasts as a relevant information source is not new.
Early reflections in software engineering already described podcasts as an emerging medium that provides expert interviews and field updates, while noting their accessibility and ease of consumption compared to traditional outlets~\cite{Rech:2007:SEPodcasts}.
At the time, podcasts were framed primarily as an informal complement to established publication venues.
In retrospect, this assessment still seems to hold true almost 20 years later, even though what was once considered an interesting but less \enquote{respectable} format has since become a pervasive medium for professional discourse.

More recently, however, the discussion has shifted from awareness to systematic use.
Infrastructure efforts such as \emph{PodcastRE} aim to archive and make podcasts searchable and analyzable at scale, explicitly positioning them as objects of scholarly study~\cite{Morris:2019:PodcastRE}.
The development of such repositories mirrors broader movements in empirical software engineering, where curated datasets have enabled new lines of inquiry.\footnote{See, e.g., SOTorrent~\cite{Baltes:2019:SOTorrent} for studying Stack Overflow posts.}
Notably, the wish for searchable, structured access to podcast content also resonates with needs expressed by participants in our survey.

Methodological contributions have begun to address how podcasts can be integrated rigorously into research designs.
For example, \citet{Kulkov:2024:GuidePodcastsAsResource} propose a structured, multi-step approach for incorporating podcasts as qualitative data, emphasizing careful selection, sampling, analysis, and ethical reflection.
Similarly, \citet{Lundstroem:2025:EthicsOfPodcastResearch} discuss ethical considerations specific to podcast research, including distinctions between public and openly licensed content, the handling of potentially sensitive data, and the question of overt versus covert research engagement.
These contributions show that the methodological foundations for podcast-based research are only now beginning to take shape across different disciplines---creating an opportunity for the SE community to adapt and actively shape these emerging standards for its own context.

Empirical demonstrations further illustrate the potential of the medium.
\citet{Ryan:2025:PodcastBreak}, for instance, analyze publicly available podcast interviews to explore women engineers' career relaunch experiences, explicitly weighing advantages (e.g., context-rich, publicly accessible data) against limitations (e.g., lack of control over interview questions, missing follow-ups, editorial mediation).
Other studies have employed large-scale Web mining and automated transcription to detect topic trends and sentiments in podcast corpora, such as \citet{Dumbach:2024:AITrendAnalysis} in the context of AI discourse in healthcare.
Earlier work within SE on mining spoken content from developer screencasts points toward the technical feasibility of extracting analyzable data from audio sources~\cite{Moslehi:2016:MiningSpeech}, but they stop short of positioning such content as a broader empirical resource for SE research.

In sum, there is growing momentum across disciplines toward recognizing podcasts as a legitimate empirical resource.
While much of this work has emerged only in recent years---and largely outside of software engineering---it provides methodological and technical building blocks.
Our study complements these efforts by examining the SE podcast landscape itself and by investigating how SE researchers perceive podcasts as a potential source for empirical inquiry.

\section{Discussion}\label{sec:discussion}

In what follows, we discuss the main findings of our study and interpret them in light of our research questions and existing literature.
We also critically reflect on the limitations of our research approach. 

\subsection{The SE Podcast Landscape}

Our analysis of the SE podcast landscape suggests that podcasting is already an established, though still evolving, communication channel within the software engineering community.
We identified 216 SE podcasts, with a median of 34 episodes per podcast and a median episode duration of about 39 minutes.
Together, these numbers indicate a non-trivial and reasonably mature content ecosystem rather than a marginal or experimental medium.
The landscape appears to support longer discussion formats that allow more detailed explanation and reflection than short-form communication channels.
At the same time, the diversity of topics---covering technical and practical knowledge, career and social aspects, as well as industry trends---suggests that SE podcasts serve multiple informational needs within the community.

The predominance of interview-based and co-hosted formats is particularly relevant when considering podcasts as a potential research resource.
Such conversational formats naturally facilitate the exchange of experiential knowledge and practitioner perspectives, which aligns with the role of grey literature in complementing formal research evidence.
However, most podcast formats imply certain methodological constraints.
Researchers have no control over how topics are introduced, structured, or elaborated during an episode, including the specific questions asked in interview-style podcasts or the narrative choices made in monologue formats.
This lack of researcher control should therefore be considered when evaluating podcasts as empirical data sources.

Overall, the landscape suggests that SE podcasts function primarily as a complementary information channel rather than as a substitute for existing knowledge sources.
The content mix we observed (spanning technical explanation, career-related discussion, and industry perspectives) indicates that podcasts may help bridge different types of knowledge within SE.
Rather than replacing established scientific or practitioner literature, SE podcasts appear to provide an additional medium for conveying experience-based insights and community-oriented knowledge that is often difficult to capture in more formal publication formats.

\subsection{Value and Barriers to Using SE Podcasts for Research}

In the introduction, we described a day in the life of our fictional character Alex, an SE researcher who regularly listens to SE podcasts to stay up to date.
Alex's relationship with podcasts closely mirrors that of our survey respondents.
Most of them are familiar with podcasts, regularly listen in their private time, and some even do so during working hours.

However, only about half of the respondents currently consider SE podcasts relevant to their own research---either today or potentially in the future.
Compared to other information sources for their work, podcasts are ranked alongside social media, behind scientific papers, books, conference presentations, and blog posts.
This positioning reflects specific concerns, such as perceived lack of credibility, as well as practical barriers including difficulties in finding and properly citing podcast content.

At the same time, respondents expressed optimism that these challenges could be addressed.
Several concrete measures were suggested to increase the usefulness of podcasts for research.
A frequently mentioned way to enhance credibility was for podcast hosts to provide sufficient references and clear sources in episode descriptions.
Another recurring theme concerned improved visibility and discoverability.
A third common request was the availability of citable transcripts.

Regarding RQ2, how SE researchers perceive podcasts as a resource for empirical research, we would characterize the overall stance as one of cautious curiosity combined with optimism.
The community recognizes potential value, yet two prerequisites stand out: methodological guidance to ensure scientific rigor, and practical tools that enable efficient work with podcast datasets.
Both aspects are currently being explored and developed across disciplines, creating an opportunity for the SE community to actively contribute to and shape this conversation.

\subsection{Limitations}\label{sec:limitations}

While our study provides a snapshot of the current SE podcast landscape and researcher perceptions, some limitations should be considered when interpreting the results.
First, the identification of SE podcasts relied on a semantic inclusion criterion rather than a purely keyword-based search.
Although this approach helped avoid missing relevant podcasts that do not explicitly use predefined keywords, it introduces a degree of subjectivity.
Some podcasts mentioned by survey respondents were not retrieved through our search procedure, suggesting that alternative search strategies or broader indexing approaches might reveal additional relevant content.

Second, the survey sample is geographically concentrated, with respondents predominantly based in Europe and South America. This may partly reflect the authors' professional networks and implies that the perception results may not fully generalize across all regions.
Relatedly, the analysis focused primarily on English-language podcasts, although podcasts in other languages also exist and some listeners may prefer content in their native language.
Future work could therefore explore country- or culture-specific podcast landscapes and assess whether regional differences influence podcast usage and perception.

Overall, we do not consider these limitations severe, but they rather reflect the exploratory nature of this study.
Our results provide useful insights into the current state of SE podcasts and researcher attitudes, while also highlighting opportunities for more comprehensive analyses in future work, such as expanding search strategies or examining regional podcast ecosystems.

\section{Conclusion}\label{sec:conclusion}

Podcasts have become part of the everyday routine of many researchers---something we turn to while commuting, cooking, or unwinding at the end of the day.
Our results suggest that this familiarity has not yet translated into widespread use of podcasts as an empirical resource in software engineering.
At the same time, the SE podcast landscape is substantial, diverse in content, and rich in practitioner perspectives.
Researchers perceive clear potential but also express concrete concerns regarding credibility, discoverability, and methodological rigor.

In a community that continues to seek stronger connections between research and practice, podcasts offer an additional channel worth considering.
Practitioners are already discussing technologies, trade-offs, failures, and emerging trends in formats they control and feel comfortable with.
Similar to how professional platforms such as LinkedIn have become spaces where research and industry perspectives intersect~\cite{Wyrich:2024:LinkedIn}, podcasts may serve as another bridge by enabling researchers to engage with conversations that practitioners are already having in their own formats and spaces, rather than requiring them to enter academic venues.

In that sense, leveraging podcasts for empirical research may not begin with asking practitioners to speak differently; it may begin with listening more deliberately.
Realizing this potential requires both methodological guidance and practical infrastructure.
If the SE community chooses to engage more systematically with podcasts, i.e., by refining selection criteria, addressing ethical considerations, and developing tools for working with audio and transcripts, it can help shape emerging standards rather than merely adopt them.

\section*{Declarations}
\backmatter
\bmhead{Funding} This work has been supported by the European Union as part of ERC Advanced Grant \enquote{Brains On Code} (101052182). Marcos Kalinowski acknowledges support from the Brazilian Research Council - CNPq (Grant 312275/2023-4), Rio de Janeiro's Research Agency - FAPERJ (Grant E-26/204.256/2024), and the Kunumi Institute. 

\bmhead{Ethical approval} Not applicable.

\bmhead{Informed consent} Informed consent exists and is available in its exact wording as part of the replication package.

\bmhead{Author Contributions} Marvin Wyrich, Marcos Kalinoswski, Adolfo Neto, and Sven Apel jointly conceptualized and designed the study. Data collection and analysis were conducted collaboratively, with all authors contributing to the classification and interpretation of the collected data. The first author led the writing of the original draft, designed the figures, and performed the majority of the final data synthesis. All co-authors contributed to reviewing, revising, and refining the manuscript.

\bmhead{Data Availability Statement}
For transparency and reproducibility, we share our study artifacts online: \url{https://doi.org/10.5281/zenodo.18962642}.
This includes a tabular overview of the sampled podcasts and their extracted attributes.
It also includes the questionnaire and anonymized responses.

\bmhead{Conflict of Interest} Not applicable.

\bmhead{Clinical Trial Number} Not applicable.

\bmhead{Acknowledgements}
We would like to thank everyone who took part in our survey.

\bibliography{main}

\end{document}